*Article*

# Linear and Nonlinear Optical Properties of SiO₂/TiO₂ Heterostructures Grown by Plasma Enhanced Atomic Layer Deposition

Jinsong Liu [1,†], Martin Mičulka [1,3,†], Raihan Rafi [1], Sebastian Beer [1], Denys Sevriukov [1,2], Stefan Nolte [1,2], Sven Schröder [2], Andreas Tünnermann [1,2], Isabelle Staude [1,4], and Adriana Szeghalmi [1,2,*]

[1] Friedrich Schiller University Jena, Institute of Applied Physics, Abbe Center of Photonics, Albert-Einstein-Str. 15, 07745 Jena, Germany
[2] Fraunhofer Institute for Applied Optics and Precision Engineering, Albert-Einstein Str. 07, 07745 Jena, Germany
[3] The Australian National University, Department of Quantum Science and Technology, Research School of Physics and Engineering, Canberra, 2601 ACT, Australia
[4] Friedrich Schiller University Jena, Institute of Solid State Physics, Max-Wien-Platz 1, 07743 Jena, Germany

\* Correspondence: adriana.szeghalmi@iof.fraunhofer.de; Tel.: +49-3641-807-40
† These authors contributed equally to this work.

**Abstract:** Second harmonic (SH) radiation can only be generated in non-centrosymmetric bulk crystals under the electric-dipole approximation. Nonlinear thin films made from bulk crystals are technologically challenging because of complex and high temperature fabrication processes. In this work, heterostructures made of amorphous materials SiO₂ and TiO₂ were prepared by a CMOS-compatible technique named plasma enhanced atomic layer deposition (PEALD) with deposition temperature at 100 °C. By using the uniaxial dispersion model, we characterized the form-birefringence properties, which can enable the phase matching condition in waveguides or other nonlinear optical applications. By applying a fringe-based technique, we determined the largest diagonal component of the effective second-order bulk susceptibility $\chi^{(2)}_{zzz}$ = 1.30±0.13 pm/V at a wavelength of 1032 nm. Noteworthy, we observed strong SH signals from two-component nanolaminates, which are several orders of magnitude larger than from single layers. The SH signals from our samples only require the broken inversion symmetry at the interface. Here optical properties of nanocomposites can be precisely tuned by the promising PEALD technology.

## 1. Introduction

Shortly after the invention of pulsed ruby laser, second harmonic generation (SHG) was observed by Franken et al. [1]. In general, nonlinear optical phenomena are described as the power series expansion of the electric field strength [2]. SHG is the nonlinear process where the induced polarization in the matter scales quadratically with the electric field. In this process, two incident photons at fundamental frequency are annihilated and one photon at the doubled frequency is generated [3]. Under the electric-dipole approximation, the SHG process is governed by a third-rank tensor [4,5]. Thus, the material symmetry can have a large influence on the SHG process. For example, SHG is forbidden in centrosymmetric bulk materials due to symmetry restrictions. When a laser beam interacts with a centrosymmetric bulk material, the bulk itself has no contributions to the SHG on the macroscopic scale under the electric dipole approximation. However, a SHG signal can arise from a very thin layer at the interface of the centrosymmetric material, as the inversion symmetry is broken in that microscopic area [6–9].

Metamaterials can introduce artificial properties which are not usually present in natural materials, such as negative refraction [10–12], hyperbolic dispersion [13,14], chirality in nanophononics [15,16], optical cloaking [17,18] and nonlinear enhancement [19–21]. Nonlinear metamaterials have been extensively studied. Nonlinear optical excitation can arise either from the intrinsic nonlinear response of the material or from artificially engineered nanostructures [22,23]. A general artificial method is to grow heterostructures. For example, ABC-type nanolaminates with three different amorphous materials can produce SHG signals which is not possible in the bulk version of the three constituents [24]. The strategy of the ABC-type nanolaminates is to utilize the heterostructures to integrate second-order interface nonlinearities. The integration of these interfaces can lead to an effective bulk nonlinearity for the overall structure [25]. The ABC-type nanolaminate can not only break the inversion symmetry at the interface, but it can also break the inversion symmetry globally. In comparison, the AB-type nanolaminate only breaks the inversion symmetry at the interface between two materials, the entire structure is still symmetric [26]. Thus, the SHG signal in the AB-type nanolaminates should be theoretically negligible. In this case, the surface susceptibilities of two interfaces would have the same absolute value but opposite sign (see Figure 1). However, interface roughness and intermixing can lead to a slight difference of atomic arrangement at the AB (blue) and BA (green) sequences. This slight difference and a small phase shift $\Delta\phi$ between SH waves generated at adjacent interfaces can lead to non-negligible SH signals. F. Abtahi et al. reported that SH waves could be generated in AB-type stacks [27]. Additionally, electric quadruple components may play a significant role, as the electric field is altered at the interface between the low and high refractive index materials.

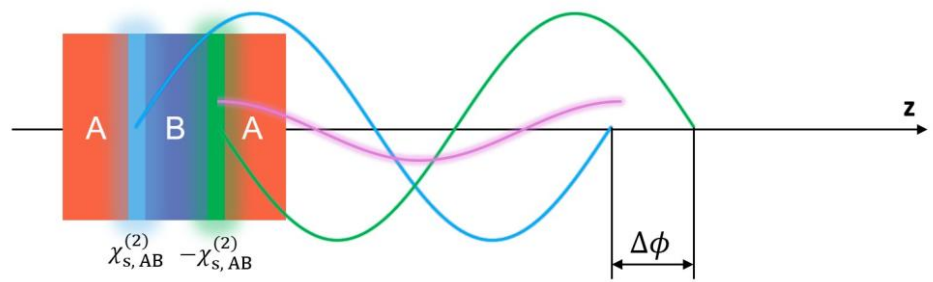

**Figure 1.** The symmetry property of AB-type nanolaminates. A small phase shift $\Delta\phi$ between the SH harmonics generated from adjacent interfaces (blue and green) leads to an incomplete destructive interference (purple).

In addition, the ABC-type nanolaminates also possess birefringence with amorphous isotropic layers [28]. The form-birefringence originates from the difference in boundary conditions of electromagnetic waves with respect to the parallel and perpendicular directions [29,30]. The very thin layers in the nanolaminates make the deposition period thickness much smaller than the wavelength of the light. Thus, the whole multilayer structure can be regarded as an effective medium with in-plane isotropy. With the rapid development of the nanoscale fabrication processes, form-birefringence is also applied to control the optical phase in optical and photonic devices [31,32].

Atomic layer deposition (ALD) has received increasing interest for optical applications because this technique has the ability to precisely control film thickness at Ångstrom level [33]. This powerful deposition technique enables the fabrication of metamaterials with unique and tailored properties, opening new possibilities in various technological applications. ALD has been used to deposit oxides [34–36], nitride [37], and thin metal films [38,39]. ALD has also been used in many applications, such as lithium batteries [40], anti-reflection coating [41–44] and metasurfaces [45,46]. In conventional ALD processes, thermal energy is utilized to activate reactions on the surface. However, high temperatures can cause thermal decomposition and break the bonds of the precursor which leads to lower growth per cycle (GPC) and more impurities [47]. In PEALD processes, the plasma radicals which can significantly reduce the deposition temperature [48]. Therefore, this process can be realized in a lower temperature range with fewer impurities than thermal ALD. In this work, the PEALD is utilized to realize the deposition of alternating thin dielectric layers.

This work aims to investigate the linear and nonlinear optical properties of the AB-type nanolaminates made of the two amorphous centrosymmetric materials $SiO_2$ and $TiO_2$ (see Figure 2). The two-component heterostructures were analyzed to understand the tunability of their optical and structural properties as the individual thicknesses change from approximately 1 nm to 0.5 nm. The SH signal generated by a loosely focused fs-pulsed laser was measured as function of the angle of incidence. From these measurements the $\chi^{(2)}$ parameters $\chi^{(2)}_{zzz}$ and $\chi^{(2)}_{zxx} + 2\chi^{(2)}_{xxz}$ were determined. The bulk like SHG properties were evaluated by power dependent measurements for films with the same composition but increasing thickness. AB heterostructures were finally compared to other nonlinear thin films and bulk nonlinear crystals.

## 2. Materials and Methods

*2.1 Deposition of $SiO_2$/$TiO_2$ Heterostructures*

The $SiO_2$/$TiO_2$ (ST) heterostructures were deposited on silicon and fused silica (FS) substrates using a SILAYO-ICP330 PEALD tool (Sentech Instruments GmbH, Berlin, Germany) [49]. To prepare the heterostructures, bis[diethylamino]silane (BDEAS) and oxygen ($O_2$) were used as precursors for $SiO_2$, tetrakis(dimethylamino)titanium (TMDAT) and $O_2$ were used as precursors for $TiO_2$, respectively. The growth of individual layers of $SiO_2$ and $TiO_2$ depend on several parameters, such as precursors' pulse duration, purge duration, deposition temperature and carrier gas flow [50,51]. The deposition temperature for all processes in this work was maintained at 100 °C. Nitrogen ($N_2$) was used both as precursor carrier gas and purge gas. The corresponding PEALD process parameters of the single layers of heterostructures are summarized in Table 1. All the parameters were previously optimized to ensure good uniformity and reproducibility.

**Table 1.** PEALD process parameters for $SiO_2$/$TiO_2$ heterostructures.

| Parameters | $SiO_2$ | $TiO_2$ |
| --- | --- | --- |
| Deposition temperature (°C) | 100 | 100 |
| Precursor (reactant) | BDEAS: $SiH_2[N(CH_2CH_3)_2]$ | TDMAT: $Ti[N(CH_3)_2]_4$ |
| Pulse (ms) | 320 | 3120 |
| Purge (ms) | 5000 | 5000 |
| Plasma gas (co-reactant) | $O_2$ | $O_2$ |
| Pulse (ms) | 3000 | 5000 |
| Purge (ms) | 2000 | 5000 |
| Growth per cycle (Å/cycle) | 1.2 | 0.7 |

*2.2 Theoretical Model and Characterization Method of Linear Optical Properties*

The AB-type heterostructures exhibit so-called form-birefringence with amorphous isotropic layers [52]. The form-birefringence is an artificial property which originates from anisotropy at a scale larger than the atomic level, but much smaller than the wavelength of light. For a two-component composite, $\varepsilon_1$ and $\varepsilon_2$ are the dielectric functions of two constituents, respectively. With the effective medium approximation, the effective refractive index of the multilayer structure can be described by equation (1) and (2).

$$n_o^2 = \varepsilon_\parallel = f_1\varepsilon_1 + f_2\varepsilon_2 \tag{1}$$

$$n_e^2 = \varepsilon_\perp = \frac{\varepsilon_1\varepsilon_2}{f_1\varepsilon_2 + f_2\varepsilon_1} \tag{2}$$

$$n_o^2 - n_e^2 = \frac{f_1 f_2 (\varepsilon_1 - \varepsilon_2)^2}{f_1\varepsilon_2 + f_2\varepsilon_1} \geq 0 \tag{3}$$

where $f_1$ and $f_2$ are volume fractions of each component. If both dielectric functions are positive, this effective medium behaves like a negative uniaxial crystal as proven in equation (3). The illustration of the multilayer and corresponding wave vector surfaces is shown in Figure 2. If one of the dielectric functions is negative, the effective medium exhibits hyperbolic dispersion relation, which is also known as hyperbolic metamaterials [53,54].

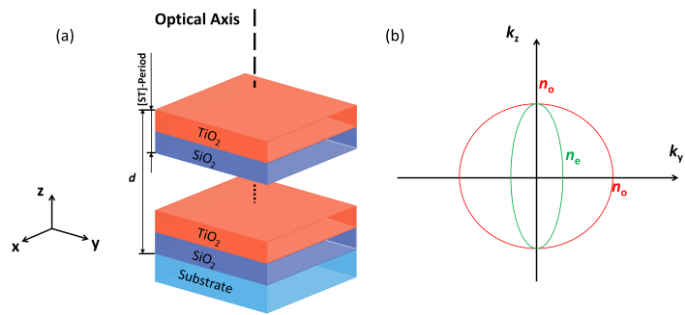

**Figure 2.** Illustration of the heterostructures and the optical axis of the uniaxial effective medium.

To extract the linear optical properties, ellipsometry and spectrophotometry were applied. Refractive indices and thickness of the heterostructures were determined by ellipsometry. The heterostructures grown on silicon substrates were measured using a SE850 DUV ellipsometer (Sentech Instruments GmbH, Berlin, Germany). A uniaxial Sellmeier dispersion model was implemented to extract physical parameters from the measured ellipsometric parameters.

Spectrophotometry was used to determine the transmittance (*T*) and reflectance (*R*) of the samples on the FS substrate. In this work, the commercial spectrophotometer 'Lamda 850' (Lambda 850, PerkinElmer Inc.,Waltham, MA, USA) was used and measurements were performed at an angle of incidence of 6° in the wavelength range from 400 nm to 1200 nm. The optical losses (*OL*) were calculated by the following expression [55]: *OL = 1 - T - R*.

X-ray reflectivity (XRR) was also implemented to estimate the thickness, surface/interface roughness, and layer densities. The XRR instruments (Bruker AXS, Karlsruhe, Germany) uses a monochromatic X-ray beam (Cu-K$\alpha$ radiation at $\lambda$ = 0.154 nm) for such measurements at a grazing incident angle from 0° to 8°. The measured spectra were analyzed by using the Brucker Leptos 7 software.

*2.3 Theoretical Model and Characterization Method of Nonlinear Optical Properties*

Following the discovery of SHG, the Maker fringe technique laid the foundation for determining the second-order susceptibility of nonlinear bulk crystals [56]. The fringes occur as the phase matching condition is not satisfied in the SHG process and the SH waves generated at different positions in the crystal can interfere constructively or destructively. For traditional nonlinear materials, Maker fringes are visible only if the thickness of the nonlinear material is significantly larger than the coherent built-up length [57]. For the nanolaminates with total thickness much smaller than the coherent built-up length, the Maker fringes are invisible. A. Hermans et al. [58] developed an effective $\chi^{(2)}$ determination method which is similar to the Maker fringe technique. In this technique, a fixed polarization of the fundamental beam is utilized, and the SH power is recorded while rotating the sample. With this method, a fringe can also be observed, and this fringe can be used to extract the effective bulk $\chi^{(2)}$ of the heterostructures. The illustration of the SHG from the AB-type heterostructures on a fused silica (FS) substrate is shown in Figure 3. The heterostructure is represented by the effective medium approximation (EMA).

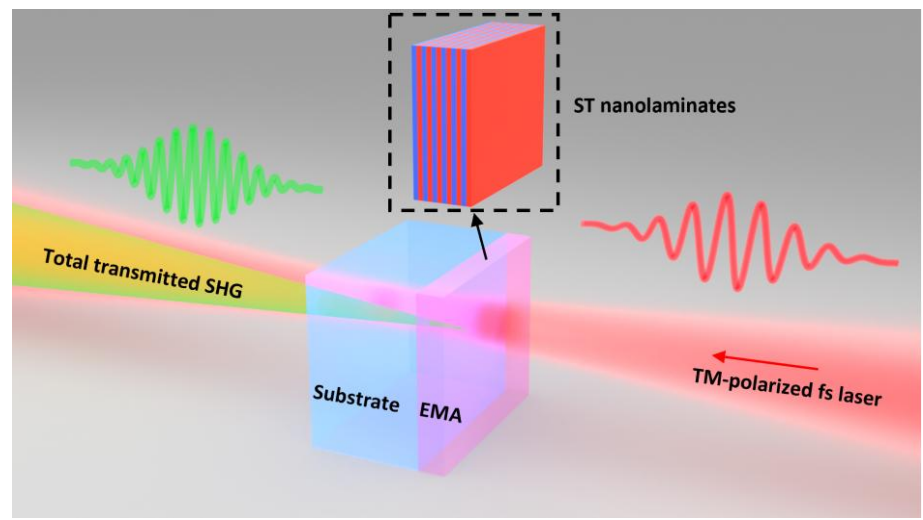

**Figure 3.** Schematic diagram of SHG in AB-type heterostructures.

In the theoretical model, the heterostructure is regarded as one homogeneous effective medium on the substrate. Furthermore, the EMA layer is assumed to have zero thickness since the total thickness is much smaller than the light wavelength and coherent built-up length. There are two contributions to the total SH signal: SH waves generated at the front interface between air and the AB film. And SH waves generated at the backside between FS substrate and air. The interference of both contributions leads to the fringe pattern. Thus, the nonlinearities are described by two effective surface second-order susceptibility tensors $\chi^i_{s,\text{eff}}$ ($i$ = AB/FS). All the interface/surface nonlinearities of the heterostructures are integrated into the first effective surface susceptibility $\chi^{\text{AB}}_{s,\text{eff}}$. Then, the first effective surface susceptibility divided by the total thickness $d$ of the heterostructures, resulting in a conversion to the effective bulk susceptibility for the overall AB heterostructure. The multiple reflections inside the heterostructures and the substrate are neglected since in the effective bulk $\chi^{(2)}$ determination process, we only consider the angle of incidence (AOI) smaller than 70°. The refractive index of the AB film is the average of the ordinary and extraordinary refractive indices as the nanolaminate is assumed to be homogeneous and optically isotropic to simplify the fitting algorithm.

The AB-type nanolaminates belong to the $C_{\infty,v}$ symmetry group where the non-zero components of the second-order susceptibility tensor components are $\chi^{(2)}_{xxz} = \chi^{(2)}_{yyz} =$ $\chi^{(2)}_{xzx} = \chi^{(2)}_{yzy}, \chi^{(2)}_{zxx} = \chi^{(2)}_{zyy}$ and $\chi^{(2)}_{zzz}$. Here, $x$ and $y$ are the in-plane directions, and $z$ is along the direction perpendicular to the plate. The total transmitted TM-/TE-polarized electric field of the layered structure with in-plane isotropy can be expressed in the form: $E_{2\omega,p} = fE^2_{\omega,p} + gE^2_{\omega,s}$, $E_{2\omega,s} = hE_{\omega,s}E_{\omega,p}$, where p and s represent TM-polarized and TE-polarized light respectively. The quantities $f$, $g$ and $h$ are rigorously deduced from the theory [28]. TM polarized light was selected as the input fundamental beam polarization state. Thus, for a monochromatic TM polarized fundamental beam $E_{\omega,\text{in}}$, the total transmitted second harmonic electric field from AB type nanolaminate on the FS substrate is given by [58].

$$E_{2\omega,\text{total}} = E_{2\omega,\text{front}} + E_{2\omega,\text{back}}$$
$$= -j\frac{\omega}{2c}t^2_{\text{air,AB}}E^2_{\omega,\text{in}}\left[\frac{T_{\text{AB,FS}}T_{\text{FS,air}}}{N_{\text{AB}}\cos(\Theta_{\text{AB}})}\chi^{\text{AB}}_{s,\text{eff}}\exp\left(\frac{-j2\omega N_{\text{FS}}\cos(\Theta_{\text{FS}})L_{\text{FS}}}{c}\right)\right.$$
$$\left. - \frac{t^2_{\text{AB,FS}}T_{\text{FS,air}}}{N_{\text{FS}}\cos(\Theta_{\text{FS}})}\chi^{\text{FS}}_{s,\text{eff}}\exp\left(\frac{-2j\omega n_{\text{FS}}\cos(\theta_{\text{FS}})L_{\text{FS}}}{c}\right)\right]$$

(4)

with $\chi^i_{s,\text{eff}}$ the effective surface second-order susceptibility approximately expressed by

$$\chi^2_{s,\text{eff}} \approx \left(\chi^i_{s,zxx} + 2\chi^i_{s,xxz}\right)\sin(\Theta_i)cos^2(\theta_i) + \chi^i_{s,zzz}\sin^2(\theta_i)\sin(\Theta_i) \quad (5)$$

In these expressions, the parameters written in lower case letters (capital letters) correspond to the fundamental frequency $\omega$ (doubled frequency $2\omega$). The Fresnel coefficients for p-polarized light propagating from medium $i$ to medium $j$ are $t_{i,j}$ and $T_{i,j}$. $n_i$ ($N_i$) and $\theta_i$ ($\Theta_i$) are the refractive indices and the propagation angles with respect to the surface normal of medium $i$. $L_{\text{FS}}$ refers to the thickness of the fused silica substrate and $c$ is the speed of light. When the $\Theta_i \approx \theta_i$ is satisfied, the approximation in equation (5) can be made. However, this approximation also leads to a mixed-up determination between $\chi^{\text{AB}}_{s,zxx}$ and $\chi^{\text{AB}}_{s,xxz}$. Thus, a combined value $A^{\text{AB}}_{zx} = \chi^{\text{AB}}_{s,zxx} + 2\chi^{\text{AB}}_{s,xxz}$ is used as a fitting parameter in the effective bulk $\chi^{(2)}$ determination process.

To determine the effective bulk $\chi^{(2)}$, the SH power from the sample needs to be considered. With the total transmitted second harmonic electric field in equation (4), we can evaluate the second harmonic power by [58]

$$P_{2\omega} = K_1|E_{2\omega,\text{front}} + E_{2\omega,\text{back}}|^2 \quad (6)$$

where $K_1$ is a calibration factor. As we are using a femtosecond (fs) laser source, the temporal walk-off time should be included in the theoretical model of SH power. The temporal walk-off time occurs since the FS substrate delays the doubled frequency ($2\omega$) pulse generated at the front interface with relative to the fundamental frequency ($\omega$) pulse. Thus, there will be a time delay between SH pulses generated at front and back interfaces. With the temporal walk-off time, the theoretical SH power can be evaluated by [58]

$$P_{2\omega} = K_2\int_{-\infty}^{+\infty}\left|E_{2\omega,\text{front}}\text{sech}^2\left(\frac{t}{\frac{\Delta t}{2\ln(1+\sqrt{2})}}\right) + E_{2\omega,\text{back}}\text{sech}^2\left(\frac{t+t_{\text{walk-off}}}{\frac{\Delta t}{2\ln(1+\sqrt{2})}}\right)\right|^2 dt \quad (7)$$

where $\Delta t$ is the FWHM pulse duration, $t_{\text{walk-off}}$ is the angle-dependent walk-off time and $K_2$ is also a calibration factor. The walk-off time can be expressed as $t_{\text{walk-off}} = t_{\text{walk-off,0}}/\cos(\theta_{\text{FS}})$ with $t_{\text{walk-off,0}}$ is the walk-off time for normal incidence. The $t_{\text{walk-off,0}}$ can be calibrated by applying equation (6) and equation (7) to the bare ~1mm FS substrate. With the theoretical model, we can measure the SH power from the sample and implement the fitting process to determine unknown parameters. The fitting parameters for SiO$_2$/TiO$_2$ nanolaminates are $K_1$, $L_{FS}$, $K_2$ and $\chi_{s,eff}^{AB}$. The surface susceptibilities of the FS substrate were found from the literature with $\chi_{s,zzz}^{FS}$=59, $\chi_{s,xxz}^{FS}$=7.9 and $\chi_{s,zxx}^{FS}$=3.8 in units of 10$^{-22}$ m$^2$/V at 1064 nm [59]. And these susceptibilities can be converted to the values at a wavelength of 1032 nm by applying the Miller's rule [3]. The first step for the fitting process is to fit $K_1$ and $L_{FS}$ simultaneously by using equation (6). This step aims to check the substrate thickness and to ensure that the theoretical fringes have the same peak positions as for the measured SH power as the peak positions are highly sensitive to the substrate thickness. Then with the known surface susceptibility of the FS substrate, we can fit other unknown parameters by using equation (7).

The SHG signals generated from the SiO$_2$/TiO$_2$ heterostructures are characterized by the setup shown in Figure 4. An ultrafast laser (Satsuma by Amplitude) with central wavelength of 1032 nm, pulse duration of 270 fs and a repetition rate of 200 kHz is used as the light source. The input power is adjusted via a half wave plate (WPH05M-1030 by Thorlabs) and a polarizer (GT10-B by Thorlabs). The second half wave plate adjusts the input polarization of the beam. The beam is focused on the sample using a plano-convex 400 mm focal length lens (LA1172-B by Thorlabs). The beam spot diameter (1/$e^2$) on the sample is around 123±3 μm. Behind the lens, a long pass filter (FELH0900 by Thorlabs) suppresses the spurious generated SH by the optical components preceding the sample. The short pass filter (FESH0800 by Thorlabs) behind the sample is used to separate the transmitted fundamental light from weak SHG signal. This filter is highly transmissive from 500 nm to 789 nm. A 520 nm bandpass filter (FBH515-10 by Thorlabs) with a FWHM of 40 nm is used to remove the ambient light and higher order harmonics. After passing through the bandpass filter, a 100 mm bi-convex lens (LA1608-A by Thorlabs) is used to image the sample onto a camera for detecting the SHG signal (CS165MU1/M by Thorlabs). The integrated pixel value over the SHG spot is calibrated to an optical power-meter so that the average power of the SHG beam can be determined. The calibration factor of camera and calculation details were reported previously [60].

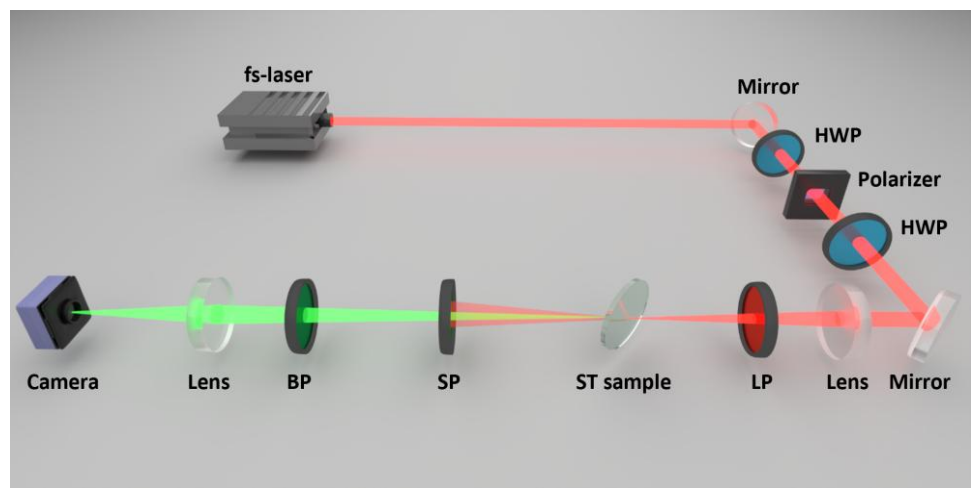

**Figure 4.** Schematic diagram of the SHG measurement setup [58]. HWP stands for half wave plate; LP, SP and BP correspond to long pass, short pass and bandpass filter, respectively.

The angular and power dependency of the SHG signals emitted by the sample are investigated. The straight transmitted SHG light is measured. For the angular dependency, the SHG is measured over a range of AOI with a fixed laser power. The AOI corresponding to the maximum SHG signal is determined, this AOI was then kept fixed for the power-dependence measurements. The fundamental laser power is gradually increased from 10 mW to 858 mW while recording the SHG signal.

## 3. Results

*3.1 Linear Optical Properties of SiO$_2$/TiO$_2$ Nanolaminates*

3.1.1. Refractive index and film thickness of SiO$_2$/TiO$_2$ nanolaminates

To analyze the linear optical properties of SiO$_2$/TiO$_2$ nanolaminates with ellipsometry, an appropriate dispersion model is applied. For the heterostructures, due to the form-birefringence property, the effective medium can be regarded as a homogeneous uniaxial medium with optical axis along the surface normal. Thus, an uniaxial Sellmeier model has been used to analyze all the heterostructures. $\Psi$ and $\Delta$ from ellipsometry measurement could be fitted well as shown in Figure 5. The uniaxial dispersion model is based on the Sellmeier dispersion model which utilizes two oscillator layers to characterize the ordinary ($n_o$) and extraordinary ($n_e$) refractive inices, respectively. During the fitting procedure, the physical parameters such as film thickness, ordinary and extraordinary refractive indices were fitted.

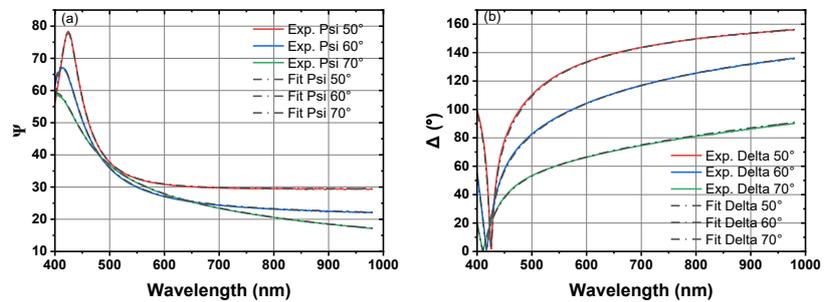

**Figure 5.** Measured ellipsometric parameters and fitting by uniaxial anisotropic model for the sample SiO$_2$:TiO$_2$ = [8:12]*30: (a) $\Psi$ and (b) $\Delta$.

Figure 6 shows fitted curves for $n_o$ and $n_e$ for the sample SiO$_2$:TiO$_2$ = [8:12]*30, 8 and 12 correspond to the ALD cycles of SiO$_2$ and TiO$_2$ respectively in one period and 30 is the number of supercycles. The solid line represents the ordinary refractive index, and the dashed line denotes the extraordinary refractive index. According to the above equations (1), (2) and (3), the effective medium should behave like a negative uniaxial crystal, and the fitted refractive indices are consistent with the theory.

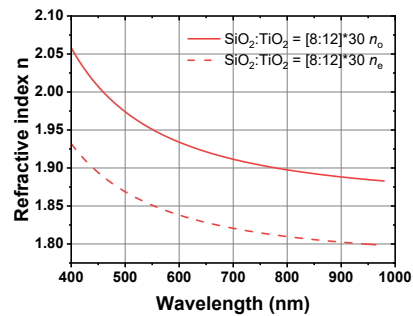

**Figure 6.** Fitting example for the sample SiO$_2$:TiO$_2$ = [8:12]*30. Solid line shows the ordinary refractive index and dashed line shows the extraordinary refractive index.

The uniaxial anisotropic dispersion model was subsequently used to fit all other SiO$_2$:TiO$_2$ compositions. The information of all the ST samples is given in Table 2. We first analyzed all the samples which have the same target total thickness, varying [ST]-period thicknesses, but the same volume fraction of each component. The fitted refractive indices for samples with a target thickness of 54 nm are shown in Figure 7. It should be noticed that when the period thickness changes, $n_o$ changes slightly, while $n_e$ changes more noticeably. The sample with thinnest period exhibits the smallest difference between $n_o$ and $n_e$. When the period thickness becomes thicker, the $\Delta n = n_o - n_e$ also becomes larger. Although these four samples have the same target total thickness, the fitted total thicknesses are different. For the deposition period [4:6], the fitted total thickness is 52 nm which is smaller than the target thickness. The target individual layer thickness of this sample is smaller than 0.5 nm, the intermixing at the interface could play an important role since the interface roughness is comparable to the individual layer thickness. There is also a change in the growth of the materials in initial 1-2 ALD cycles. For the deposition period [10:15], the fitted total thickness is 54.8 nm. The difference between the total thicknesses of different periods [4:6] and [10:15] of 2.8 nm is however small. Both samples have exactly the same number of $SiO_2$ (240) and $TiO_2$ (360) deposition cycles. The variation in the film thickness indicates a different GPC of the oxide deposited on each other with rapid variation of the sequence than when $SiO_2$ or $TiO_2$ grows on itself. The intermixing might influence the available functional sites and hindrance due to the ligands.

**Table 2.** Deposition information and birefringence of all the ST Nanolaminates.

| Deposition cycle $SiO_2$ | Deposition cycle $TiO_2$ | Super cycles | Layer thickness(nm) $SiO_2$ | Layer thickness(nm) $TiO_2$ | Fitted thickness (nm) ±0.1 | Target thickness (nm) | $n_o$ @980nm ±0.01 | $n_e$ @980nm ±0.01 |
|---|---|---|---|---|---|---|---|---|
| 8 | 12 | 20 | 0.96 | 0.84 | 36.5 | 36 | 1.89 | 1.80 |
| 4 | 6 | 60 | 0.48 | 0.42 | 52.0 | 54 | 1.89 | 1.87 |
| 6 | 9 | 40 | 0.72 | 0.63 | 53.9 | 54 | 1.89 | 1.85 |
| 8 | 12 | 30 | 0.96 | 0.84 | 54.5 | 54 | 1.88 | 1.79 |
| 8 | 12 | 30 | 0.96 | 0.84 | 53.5* | 54 | 1.86 | 1.80 |
| 10 | 15 | 24 | 1.2 | 1.05 | 54.8 | 54 | 1.88 | 1.75 |
| 8 | 12 | 50 | 0.96 | 0.84 | 89.3 | 90 | 1.90 | 1.81 |
| 6 | 9 | 120 | 0.72 | 0.63 | 158.0 | 162 | 1.89 | 1.87 |
| 8 | 12 | 90 | 0.96 | 0.84 | 161.0 | 162 | 1.87 | 1.77 |

* Repeated deposition after 6 months

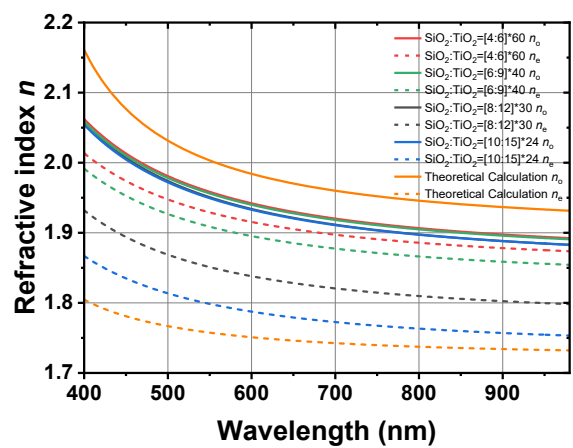

**Figure 7.** Ordinary and extraordinary refractive indices of samples with the same total thickness but different number of interfaces.

Figure 8 shows the $\Delta n$ values change with the period thickness at the wavelength of 980 nm. The theoretical calculation of the $\Delta n$ at 980 nm is 0.20. This theoretical value is higher than the $\Delta n$ values from the ellipsometry analysis. The first group of samples have the same target total thickness, i.e. varying [ST]-period thicknesses, but the same volume fraction of each component. Thus, they are expected to have the same effective refractive indices according to the theory [29]. This inconsistency can also be explained by the intermixing. These four samples have very thin individual layer thicknesses which are close to the interface roughness. Thus, the intermixing can have an obvious influence on the refractive indices.

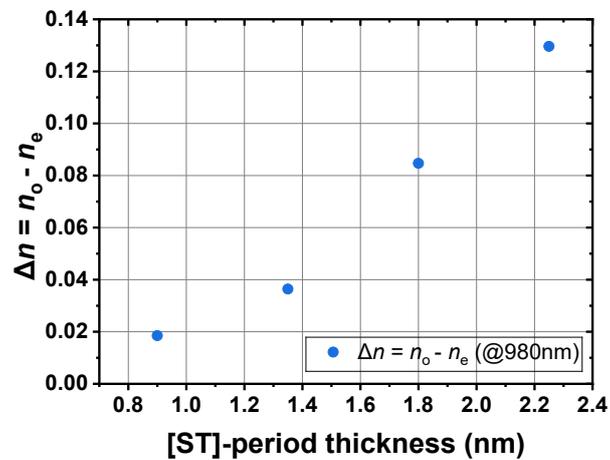

**Figure 8.** The difference between $n_o$ and $n_e$ at 980 nm wavelength for samples with the same volume fraction.

Next, samples with the same period $SiO_2:TiO_2 = [8:12]$, but different thicknesses will be discussed. The refractive indices are mainly independent on the total thickness as shown in Figure 9. The slight differences can be attributed to fitting errors, variations in film growth conditions, different thicknesses and surface roughness.

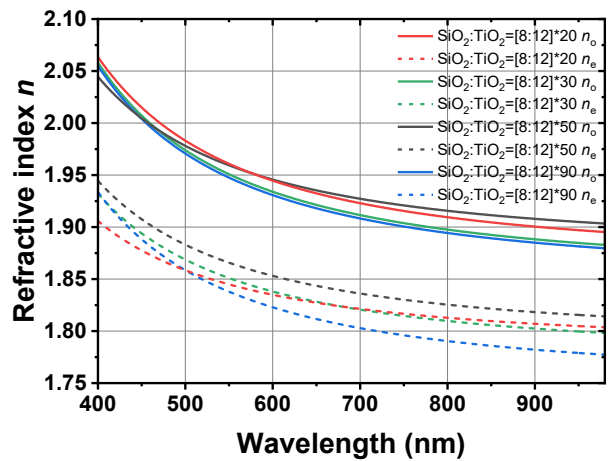

**Figure 9.** Ordinary and extraordinary refractive indices of second batch of samples.

To test the repeatability, we had two samples with the same recipe, but one of them was deposited half a year later. The difference in film thickness of 1 nm is mainly caused by the slight variations of the growth conditions and the fitted refractive indices given in supplementary material also show only slight variations.

3.1.2. Transmittance and Reflectance

Transmittance and reflectance spectra of these heterostructures at an angle of incidence of 6° are shown in Figure 10. Figure 10 (a) shows that samples with the same period but different total thicknesses possess different $T/R$ spectra as expected.

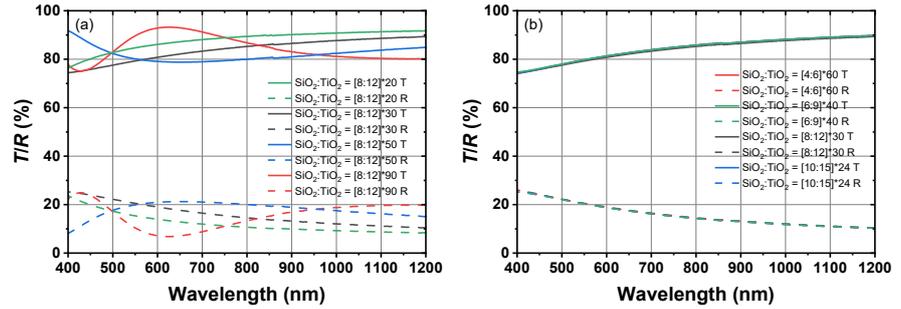

**Figure 10.** Transmittance and Reflectance: (a) *T*/*R* data of the samples with period [8:12] but different supercycles; (b) *T*/*R* data of the samples with the same total thickness but different periods.

However, Figure 10 (b) exhibits that samples with the same total thickness, but different periods have the same *T* and *R* spectra in the wavelength range 400 nm - 1200 nm at 6° of incidence. The incident angle of 6° is sufficiently small to be regarded as normal incidence. Thus, the transmittance and reflectance of these samples should be almost the same due to the similar $n_\mathrm{o}$ values. The corresponding optical losses in the wavelength range 400 nm - 1200 nm are calculated in Figure 11, which proves that our samples have low optical losses at both fundamental wavelength (FW) and second harmonic (SH) wavelength which is beneficial to obtain a good laser stability and high laser induced damage threshold.

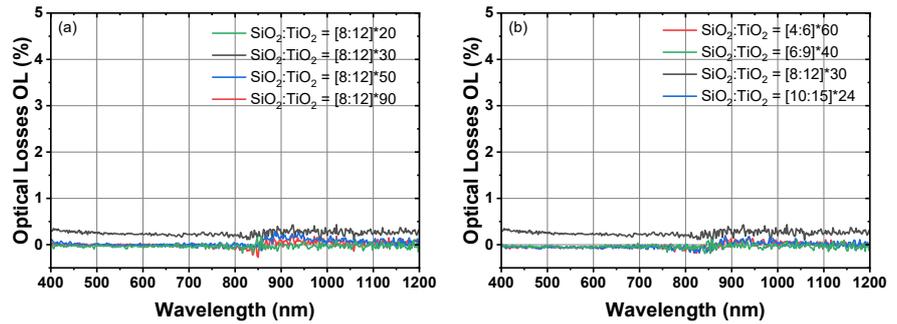

**Figure 11.** Optical losses of samples in Figure 10.

3.1.3. X-ray Reflectivity Analysis

X-ray reflectivity (XRR) is a non-destructive technique which can provide information on film thickness, intermixing at interface, layer density, and surface roughness. In this work, XRR was used to analyze two thicker samples on silicon substrates at grazing angles from 0° to 8°. The measured data and fitted fringes are shown in Figure 12.

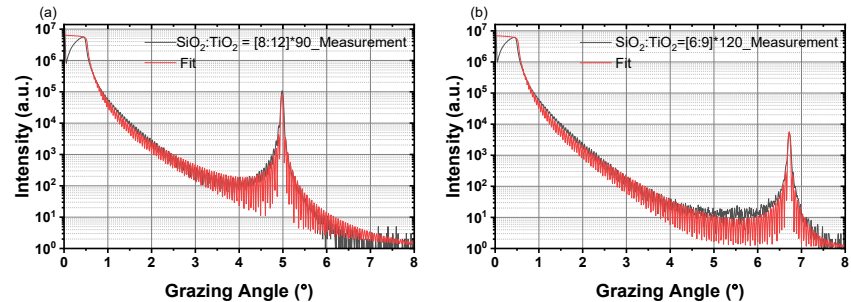

**Figure 12.** XRR comparison between $SiO_2$:$TiO_2$ = [8:12]*90 and $SiO_2$:$TiO_2$ = [6:9]*120.

The presence of the sharp Bragg peaks indicates the periodic structure. The estimated layer thicknesses of $SiO_2$ for $SiO_2$:$TiO_2$ = [6:9] and [8:12] heterostructures were 0.6 nm and 0.8 nm, respectively. The estimated layer thicknesses of $TiO_2$ were 0.7 nm and 1.0 nm. The intermixing can be estimated with the interface roughness. When the individual layer thickness becomes thinner, the interface roughness becomes higher which leads to the variations of the refractive indices in Figure 7. Table 3 summarizes film total thicknesses and interface roughness, comparing ellipsometry and XRR results. The total thicknesses from ellipsometry and XRR are almost identical.

**Table 3.** XRR data and thickness comparison with ellipsometry.

| Compositions $SiO_2$:$TiO_2$ | Interface roughness (nm) $SiO_2$ \| $TiO_2$ | Layer thickness (nm) $SiO_2$ \| $TiO_2$ | Target thickness (nm) | Total thickness (nm) XRR | Total thickness (nm) Ellipsometry |
|---|---|---|---|---|---|
| [6:9]*120 | 0.3 \| 0.6 | 0.6 \| 0.7 | 162 | 158.2 | 158.0 |
| [8:12]*90 | 0.3 \| 0.3 | 0.8 \| 1.0 | 162 | 160.5 | 161.0 |

*3.2. Nonlinear Optical Properties of $SiO_2$/$TiO_2$ nanolaminates*

3.2.1. Angular and power dependency

In the theoretical model, all the surface/interface nonlinearities are integrated into the first effective surface nonlinearity $\chi_{s,\text{eff}}^{AB}$ which can be converted to the effective bulk nonlinearity for the overall structure. To study the relationship between the number of interfaces and nonlinearities, the heterostructures have different [ST]-period thicknesses but the same volume fraction of each component was tested. Figure 13 shows the measured angular and power dependency of the average SH power. The SHG signal is clearly dependent on the number of interfaces. With more interfaces, the peak value of the SHG signal is also higher. However, for the sample [4:6]*60, which has the largest number of interfaces, the SHG signal is not the highest. Due to the fact that the individual layer thickness is only about 0.5 nm, in this case the intermixing could decrease the quality of the interfaces which leads to a reduced SH signal. From Figure 13, it is obvious that the measured average SH power is dependent on the number of interfaces. The relationship is expected to be linear dependency since the nonlinearity is proportional to the layer densities. When the layer thickness is larger than 0.5 nm, the measured SH power is consistent with this expected linear dependency (see Figure 15). As discussed before, this consistency is broken for the sample with thinnest layer thickness, the intermixing decreases the SHG signals. The power dependency is used to verify we indeed obtain the SHG, with the slope of the lines in the range of 1.97(1) to 2.07(1) as shown in Figure 14. The uncertainties with the slopes are the standard errors from linear fittings.

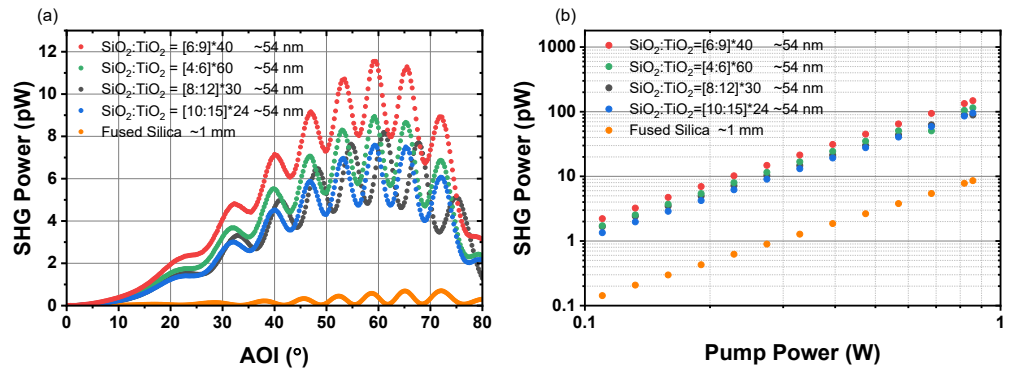

**Figure 13.** (a) Angular dependency measured at 250 mW incident power and (b) power dependency of the ST samples with the same total thickness but different periods.

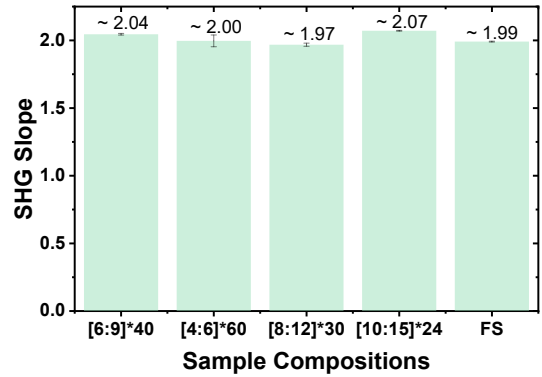

**Figure 14.** SH power slopes for the heterostructures with the same total thickness but different number of interfaces.

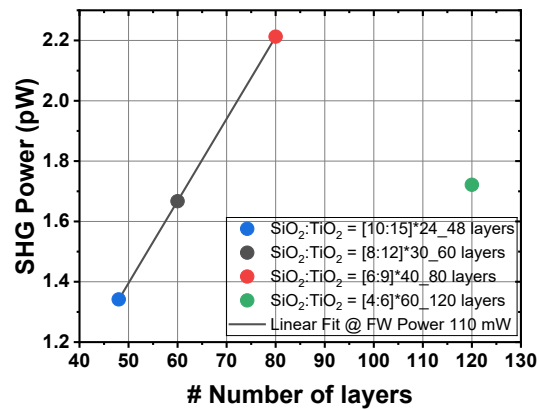

**Figure 15.** Measured average SH power vs. number of layers ($d ≈ 54$ nm). The input power at FW is 110 mW.

For bulk crystals, when the phase matching condition is satisfied, the field amplitude is linearly dependent on the propagation distance. To confirm whether the nanolaminates have similar property, the samples with the same period but different number of supercycles were also measured. The angular and power dependencies of the average SH power are displayed in Figure 16. Obviously, the SHG signals from the ST nanolaminates are much stronger than the pure fused silica substrate. The SH power clearly increases with increasing thickness. Fitting power dependency measurements yields a slope with an exponent of 1.97(1) to 2.07(1) as shown in Figure 17, verifying the generation of the second harmonic. The uncertainties with the slopes are the standard errors from linear fittings.

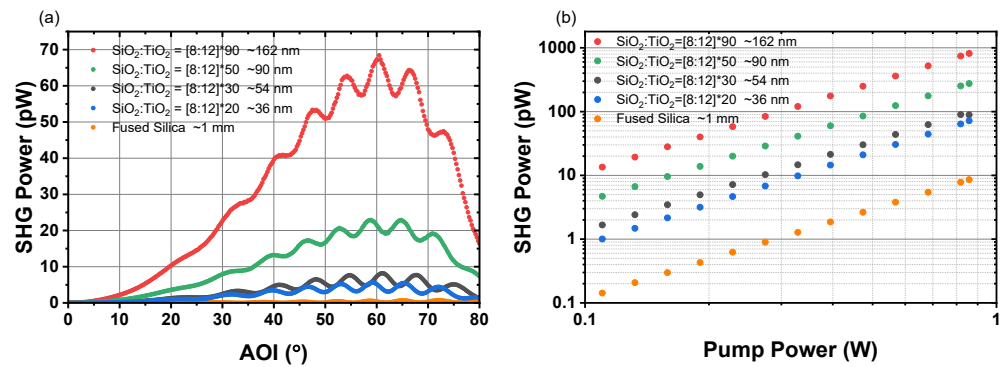

**Figure 16.** (a) Angular dependency measured at 250 mW incident power and (b) power dependency of the ST samples with different thicknesses.

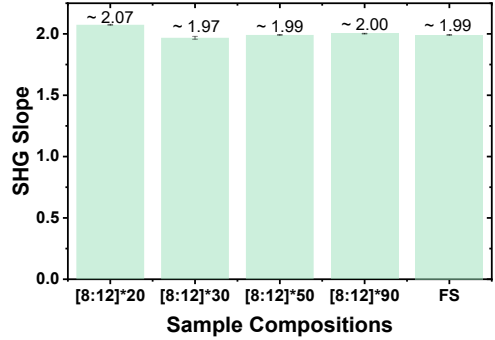

**Figure 17.** SH power slopes for the heterostructures with the same period but different supercycles.

The SH power is linearly dependent on the squared total thickness as shown in Figure 18. The total thicknesses of our samples are much smaller than the coherence built-up length and light wavelength, so that the different waves can be considered to be in phase.

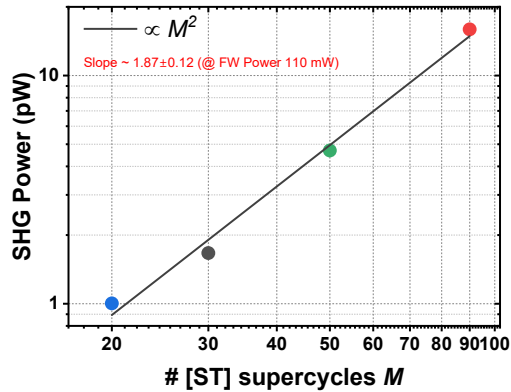

**Figure 18.** Measured average SH power vs. number of periods M=20, 30, 50 and 90 (double logarithmic scale) at a FW power of 110 mW.

The dependency of the SHG signals on the input polarization state was also tested as shown in Figure 19 for the sample SiO$_2$:TiO$_2$ = [6:9]*40. For the measurements the angle of incidence yielding the highest SH power was selected. A polarization analyzer (Rochon prism analyzer RPM10, Thorlabs) was added to analyze the polarization state of total transmitted SHG signal. The configuration is given in Figure S2. Figure 19 exhibits the fact that no TE-polarized SHG signal can be observed for both TM- and TE-polarized input FW beams. This indicates our two-component heterostructures belong to the symmetry group $C_{\infty,v}$ with in-plane isotropy. We should note that the transmitted TM-/TE-polarized electric fields with doubled frequency are expressed by $E_{2\omega,p} = fE_{\omega,p}^2 + gE_{\omega,s}^2$, $E_{2\omega,s} = hE_{\omega,s}E_{\omega,p}$. If the input FW light beam is TM-polarized, the transmitted SH wave is TM-polarized. However, if the input FW light beam is TE-polarized, the SH wave is still TM polarized due to the nonzero parameter $g$ deduced from the theory [28]. Thus, with TE-polarized input FW beams, TM-polarized SHG signals can still be observed which correspond to the weak but nonzero values in Figure 19 when the rotation angle of HWP is 45° or 135°. The measured TE-polarized SHG signals are much weaker than TM-polarized SHG signals as the diagonal component $\chi_{zzz}^{(2)}$ is much larger than other non-vanishing components [5].

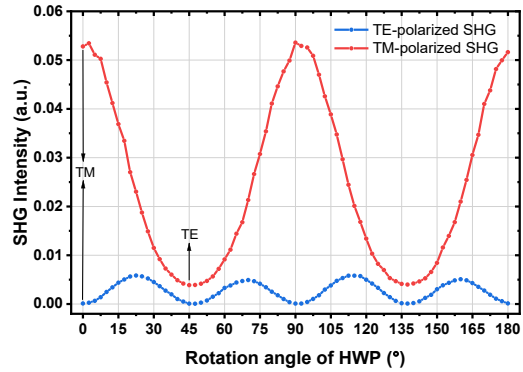

**Figure 19.** TE-polarized and TM-polarized SHG signals as a function of the linear polarization direction of the pump beam, controlled by the rotation angle of the second half wave plate. 0° and 45° of HWP correspond to the incident TM- and TE-polarization of the FW beam.

3.2.2. Effective $\chi^{(2)}$ determination

In the $\chi^{(2)}$ determination process, the first step was to measure the angle-dependent SH power generated from the heterostructures. For a sufficient resolution, the angle step was set as 0.25°. The fitting of the measured SH power was implemented by a nonlinear least-squares fitting algorithm in MATLAB. With the group index of FS (see Figure S3), the fittings for the bare FS substrate with walk-off time and without walk-off time are performed in Figure S4. The determined walk-off time was 97 fs. The peak positions of the fringes are very sensitive to the substrate thickness. Neglecting the walk-off time results in higher fringe visibility, which in turn causes the nonlinearity to be overestimated. Thus, in the fitting processes for the nanolaminates on the FS substrate, the model including walk-off time was used.

The unknown parameters e.g. substrate thickness $L_{FS}$, $K_1$, $K_2$, $A_{s,zx}^{AB}$ and $\chi_{s,zzz}^{AB}$ were fitted. The fitting results for the samples which have the same total thickness are shown in Figure 20. All the theoretical fringes match the measured data very well, thus the effective second-order bulk susceptibilities can be extracted.

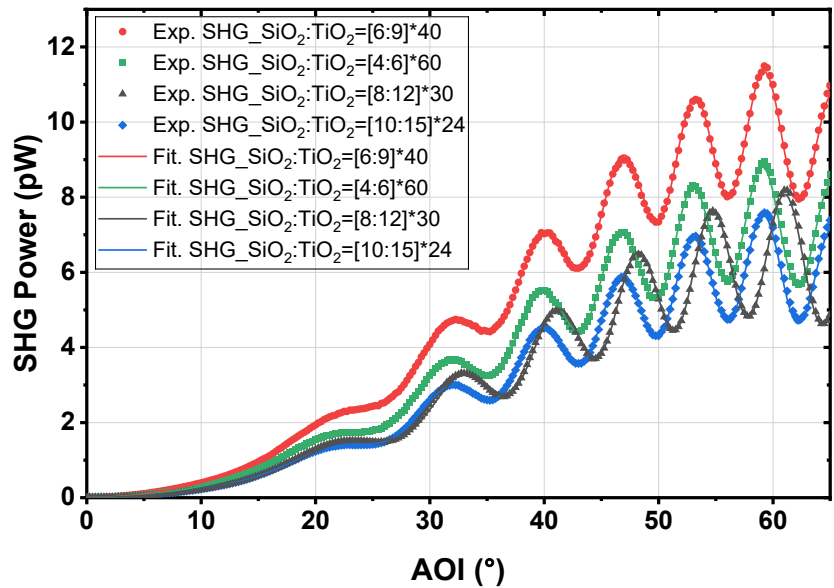

**Figure 20.** Nonlinear fitting results for the samples with the same total thickness ~54 nm: (a) $SiO_2:TiO_2$ = [6:9]*40; (b) $SiO_2:TiO_2$ = [4:6]*60; (c) $SiO_2:TiO_2$ = [8:12]*30; (d) $SiO_2:TiO_2$ = [10:15]*24.

The extracted second-order susceptibility tensor components are given in Figure 21 and Table 4. The best period composition for SHG of SiO$_2$/TiO$_2$ heterostructures is [6:9], with the main diagonal component $\chi^{(2)}_{zzz}$ = 1.30±0.13 pm/V. This result is consistent with the measured SH power in Figure 13, where the composition [6:9] of the SiO$_2$/TiO$_2$ heterostructure indeed has the highest SHG signal. The main uncertainty originates from the alignment process, where the sample orientation corresponding to the normal incidence needs to be set precisely.

**Table 4.** Fitted second-order susceptibilities

| Composition | $\chi^{(2)}_{zzz}$ (pm/V) | $\chi^{(2)}_{zxx} + 2\chi^{(2)}_{xxz}$ (pm/V) |
|---|---|---|
| SiO$_2$:TiO$_2$ = [6:9]*40 | 1.30±0.13 | 1.34±0.13 |
| SiO$_2$:TiO$_2$ = [4:6]*60 | 0.97±0.10 | 1.14±0.11 |
| SiO$_2$:TiO$_2$ = [8:12]*30 | 0.92±0.09 | 0.78±0.08 |
| SiO$_2$:TiO$_2$ = [10:15]*24 | 0.89±0.09 | 0.86±0.09 |

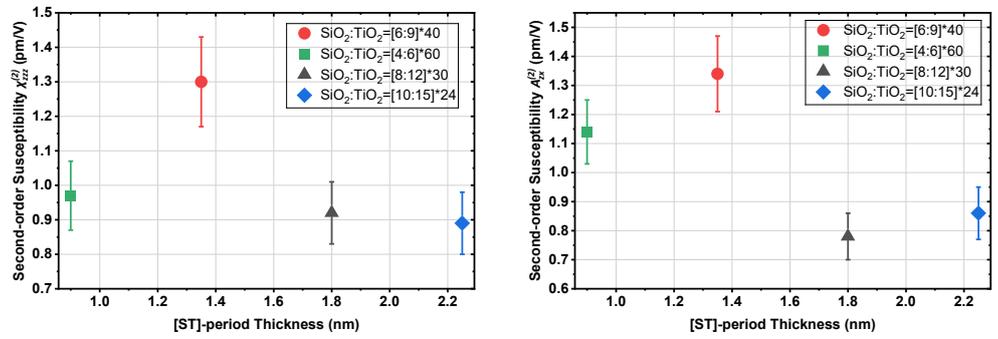

**Figure 21.** (a) Diagonal components and (b) Inseparable non-diagonal components of the second-order susceptibility tensor.

The SHG repeatability was also tested for samples with the same recipe, but one of them was deposited half a year later. The measured SH power from two samples is given in Figure S5. The $\chi^{(2)}_{zzz}$ values for the two samples were 0.92 pm/V and 0.98 pm/V, respectively.

After the best composition of ST heterostructures for SHG was found, we could compare ST heterostructures with other nonlinear materials. Table 5 summarizes the main component of second-order nonlinear susceptibility tensors of different nonlinear materials. "AB Amorphous" and "ABC Amorphous" stand for heterostructures with two or three different amorphous materials. From previous work [58], "AB Amorphous" type of heterostructures could only exhibit negligible second-order nonlinearity. Our results showed that two-component heterostructures possess non-negligible second-order nonlinearity. The origin of these effects in such "simple" compositions needs to be further investigated.

**Table 5.** Fitted second-order susceptibilities

| Grown Structure | Material | Point Group | $d$ (pm/V) | $\lambda$ (nm) | Ref. |
|---|---|---|---|---|---|
| Monocrystal | Quartz | $D_3$ | $d_{11}$ = 0.3 | 1064 | [61] |
| Monocrystal | BaB$_2$O$_4$ | $C_{3,v}$ | $d_{22}$ = 2.2 | 1064 | [61] |
| Monocrystal | BaTiO$_3$ | $C_{4,v}$ | $d_{33}$ = 6.8 | 1064 | [62] |
| Monocrystal | LiNbO$_3$ | $C_{3,v}$ | $d_{33}$ = -27 | 1064 | [63] |
| ABC Amorphous | In$_2$O$_3$/TiO$_2$/Al$_2$O$_3$ | $C_{\infty,v}$ | $d_{33}$ = 0.6 | 800 | [58] |
| ABC Amorphous | HfO$_2$/TiO$_2$/Al$_2$O$_3$ | $C_{\infty,v}$ | $d_{33}$ = 0.4 | 800 | [58] |
| AB Amorphous | SiO$_2$/TiO$_2$ | $C_{\infty,v}$ | $d_{33}$ = 0.65 | 1032 | This work |
| AB Superlattice | LaTiO$_3$/SrTiO$_3$ | $O_h$ | $d_{eff}$ = 12.15 | 800 | [26] |
| ABC Amorphous | SiO$_2$/TiO$_2$/Al$_2$O$_3$ | $C_{\infty,v}$ | $d_{33}$ = 1.0 | 1032 | [64] |

The change in the electric field and the phase shift at the interface might play a significant role to alter the interface SHG. Additionally, differences in the atomic morphology of the $SiO_2/TiO_2$ versus $TiO_2/SiO_2$ may also have influence.

In addition, H. Zhao et al. [26] studied the second harmonic generation in superlattice structures made of two centrosymmetric crystalline materials $LaTiO_3/SrTiO_3$ grown by pulsed laser deposition (PLD). The point group of the overall structure was identified quantitatively, by calculating the SHG signal intensity according to the assumed symmetry group and comparing with the measured SHG signal. The point group was determined to be $m\bar{3}m$ $(O_h)$. The effective second-order nonlinearity was determined by comparing the SHG signals from the superlattice with SHG signals from a Y-cut quartz sample. The very high effective second-order nonlinearity implied that SHG can be improved by heterostructures made of crystalline materials with high quality of the interface. For our ST heterostructures, one possible SHG enhancement method is to use an annealing process to increase the crystallinity. In our first try, samples were annealed at 600 °C. The SHG and XRR measurements are shown in Figure S6, Figure S7, Figure S8 and Table S1. For the sample $SiO_2:TiO_2 = [8:12]*90$, the Bragg peak shifted from 5° to 6° and its intensity decreased significantly after the annealing process. This indicates severe structural changes by the annealing process. The interface quality was decreased by the intermixing and only a very weak SHG signal after the annealing process was measured. For the sample $SiO_2:TiO_2 = [6:9]*120$, since it has no Bragg peak around 6°, and the interface roughness becomes as large as the individual layers, we can conclude that the layered character is lost. Thus, the SHG signal is also very weak after the annealing process. Next steps are to heat up to a lower annealing temperature and try to use other materials to further enhance the interface quality.

## 4. Conclusions

In this work, we have studied the linear and nonlinear optical properties of $SiO_2/TiO_2$ heterostructures. The heterostructures can integrate a large number of very thin layers at the atomic scale. Under the effective medium approximation, the form-birefringence in these films leads to a negative uniaxial crystal behavior. The refractive index is independent of the total thickness, however, it is tunable by combining two materials with different composition periods. When the individual layer thickness is only a few nm, the intermixing can have large influence on the refractive indices. The total thickness, ordinary and extraordinary refractive indices were determined using an uniaxial anisotropic dispersion model. The total thickness was also determined by XRR. Both characterization methods yielded reliable results.

The nonlinear optical properties of $SiO_2/TiO_2$ nanolaminates were analyzed in detail. Unlike the symmetry breaking mechanism in traditional nonlinear bulk crystals, the $SiO_2/TiO_2$ heterostructures break the symmetry only at the interface between two amorphous centrosymmetric media. In general, one should expect that the $SiO_2/TiO_2$ nanolaminates are still symmetric under inversion operations along the surface normal. Thus, a negligible SHG signal from this type of nanolaminates should be observed. Nevertheless, the global inversion symmetry is valid only when the $SiO_2/TiO_2$ and $TiO_2/SiO_2$ interfaces have the same functional and optical properties. In the real samples, they are different interfaces, and the SH waves of adjacent interfaces do not cancel. The nanolaminates consisting of very thin layers can integrate a large number of interfaces, resulting in an effective bulk nonlinearity for the overall structure. By measuring the average SH power from the samples with the same total thickness but different periods, we could verify that the SHG signal is dependent on the number of interfaces. However, when the individual layers are thinner than 0.5 nm, the SHG signal is reduced because the interface quality is decreased by the intermixing. By measuring the average SH power from the samples with the same period, but different supercycles, we could verify that the SHG signal is linearly dependent on the squared total thickness. The SHG measurements should be implemented under TM-polarized light since the main component of the second-order susceptibility tensors is the diagonal component $\chi^{(2)}_{zzz}$. While the value is lower than for typical crystals used for SHG, we obtained values comparable to more complex ABC type structures.

Overall, this work exhibits that the linear and nonlinear optical properties of the engineered heterostructures made by PEALD can be precisely controlled. Based on a low temperature (100 °C) and conformal deposition, these artificial nonlinear materials can be integrated into CMOS-compatible processes to design metamaterials and photonics integrated circuits (PICs).

**Author Contributions:** Conceptualization, A.S.; methodology, J.L., M.M., R.R., S.B., S.N. and A.S.; software, J.L., M.M. and S.B.; validation, J.L., M.M. and R.R.; formal analysis, J.L., M.M., R.R., S.B. and D.S.; investigation, J.L., M.M. and R.R.; resources, S.S.; data curation, J.L., M.M. and R.R.; writing—original draft preparation, J.L.; writing—review and editing, J.L., M.M., S.B., R.R., S.N. and A.S.; visualization, J.L.; supervision, A.T., I.S. and A.S.; project administration, S.S., A.T., I.S. and A.S.; funding acquisition, A.T., I.S. and A.S. All authors have read and agreed to the published version of the manuscript.

**Funding:** This work was supported by the Deutsche Forschungsgemeinschaft (DFG, German Research Foundation) through the International Research Training Group (IRTG) 2675 "Meta-ACTIVE" (project number 437527638) and the Collaborative Research Center (CRC) 1375, (project number 398816777) "NOA". Funding by the Federal Ministry of Research, Technology and Space, Project ID 13N16897 (Gradient) is highly acknowledged. Additional support was provided by the Fraunhofer Internal Programs under Grant No. SME 431 40-04871.

**Institutional Review Board Statement:** Not applicable

**Informed Consent Statement:** Not applicable

**Data Availability Statement:** Not applicable

**Acknowledgments:** The authors gratefully acknowledge Kristin Gerold for her technical support.

**Conflicts of Interest:** The authors declare no conflicts of interest.

## Abbreviations

The following abbreviations are used in this manuscript:

| | |
|---|---|
| PEALD | Plasma Enhanced Atomic Layer Deposition |
| ALD | Atomic Layer Deposition |
| SHG | Second Harmonic Generation |
| FS | Fused Silica |
| $SiO_2$ | Silicon Dioxide |
| $TiO_2$ | Titanium Dioxide |
| $O_2$ | Oxide |
| $N_2$ | Nitrogen |
| SH | Second Harmonic |
| XRR | X-Ray Reflectivity |
| AOI | Angle Of Incidence |
| T | Transmittance |
| R | Reflectance |
| OL | Optical Losses |
| FW | Fundamental Wavelength |
| ST | $SiO_2/TiO_2$ |
| HWP | Half Wave Plate |
| LP | Long pass |
| SP | Short pass |
| BP | Bandpass |
| GPC | Growth per cycle |

*Supplementary Information*

# Linear and Nonlinear Optical Properties of SiO$_2$/TiO$_2$ Heterostructures Grown by Plasma Enhanced Atomic Layer Deposition

**Jinsong Liu** [1,†], **Martin Mičulka** [1,3,†], **Raihan Rafi** [1], **Sebastian Beer** [1], **Denys Sevriukov** [1,2], **Stefan Nolte** [1,2], **Sven Schröder** [2], **Andreas Tünnermann** [1,2], **Isabelle Staude** [1,4], and **Adriana Szeghalmi** [1,2,*]

[1] Friedrich Schiller University Jena, Institute of Applied Physics, Abbe Center of Photonics, Albert-Einstein-Str. 15, 07745 Jena, Germany
[2] Fraunhofer Institute for Applied Optics and Precision Engineering, Albert-Einstein Str. 07, 07745 Jena, Germany
[3] The Australian National University, Department of Quantum Science and Technology, Research School of Physics and Engineering, Canberra, 2601 ACT, Australia
[4] Friedrich Schiller University Jena, Institute of Solid State Physics, Max-Wien-Platz 1, 07743 Jena, Germany

* Correspondence: adriana.szeghalmi@iof.fraunhofer.de; Tel.: +49-3641-807-40
† These authors contributed equally to this work.

Supplementary Information

To test the repeatability with respect to the linear optical properties, we made two samples with the same deposition recipe SiO$_2$:TiO$_2$ = [8:12]*30, but one of them was deposited half a year later. The difference in film thickness of 1 nm is mainly caused by slight variations of the growth conditions. The fitted refractive indices given in Figure S1 show only slight differences.

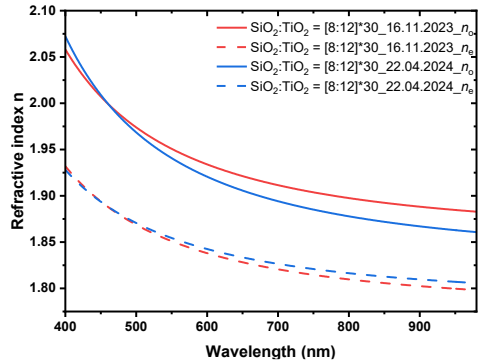

**Figure S1.** Refractive index comparison between samples with the same recipe but fabricated 6 months apart.

The SHG polarization dependency was measured with the setup in Figure 4, but an analyzer was added to verify the polarization state of the transmitted SHG signals as shown in Figure S2.

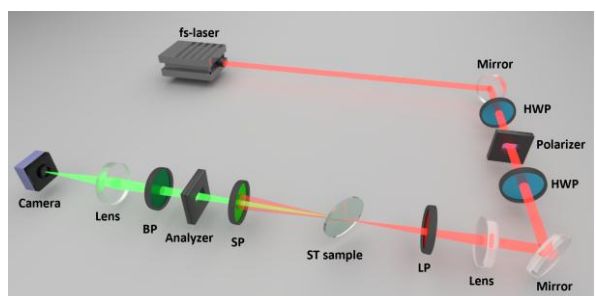

**Figure S2.** Schematic diagram of SHG polarization dependency test configuration.

The walk-off time is a small time delay between SH pulses generated at the front and the back interface. The reason for the temporal walk-off effect is that the pulse at the fundamental wavelength will travel at a different group velocity through the substrate than the SH pulse generated at the front interface. The different group velocity is caused by different group index (see Figure S3). For ~1 mm thick FS substrate, the walk-off time is estimated to be ~100 fs.

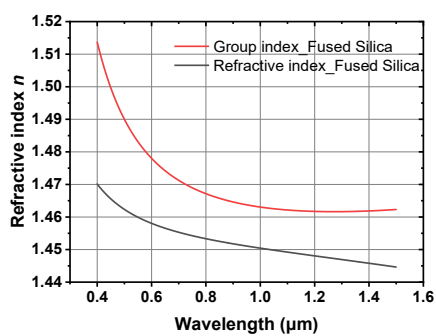

**Figure S3.** Group index and refractive index of FS substrate [1,2].

To determine an accurate walk-off time, the average SH power generated from the FS substrate was measured with the setup in Figure 4. Equation (7) without nanolaminates on the substrate is built for the FS substrate. By fitting the model with the measured average SH power, we obtained the $t_{\text{walk-off, 0}}$ = 97 fs (see Figure S4).

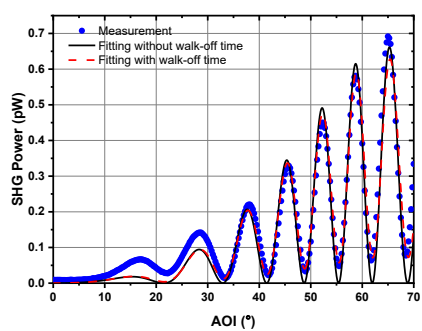

**Figure S4.** Fitting the measured SH power from the fused silica substrate with walk-off time (red dashed line) and without walk-off time (black solid line).

To test the repeatability with respect to the nonlinear properties, we made two samples with the same deposition recipe $SiO_2:TiO_2$ = [8:12]*30, but one of them was deposited half a year later. The measured SHG signals are shown in Figure S5, the measured average SH powers and corresponding fitted second order nonlinear susceptibility values are quite close.

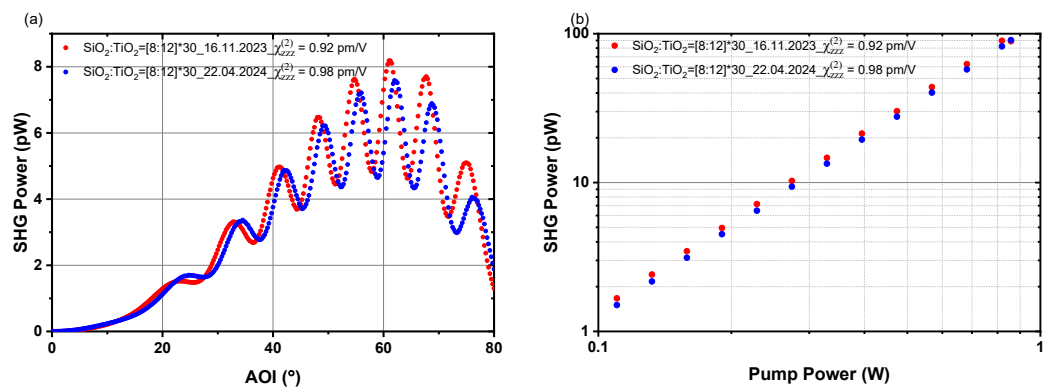

**Figure S5.** Measured SH signals (a) angular and (b) power dependency from samples with the same deposition recipe manufactured 6 months apart. The SHG slopes are 1.97(1) (red) and 2.00(1) (blue).

The SHG and XRR measurements of annealed samples $SiO_2:TiO_2$ = [8:12]*90 and [6:9]*120 are shown in Figure S6, S7 and S8. We can conclude that after the annealing process, the interface was destroyed by the intermixing, thus we could not get strong SHG signals from the annealed samples. Their SHG response decreased by 2 orders of magnitude.

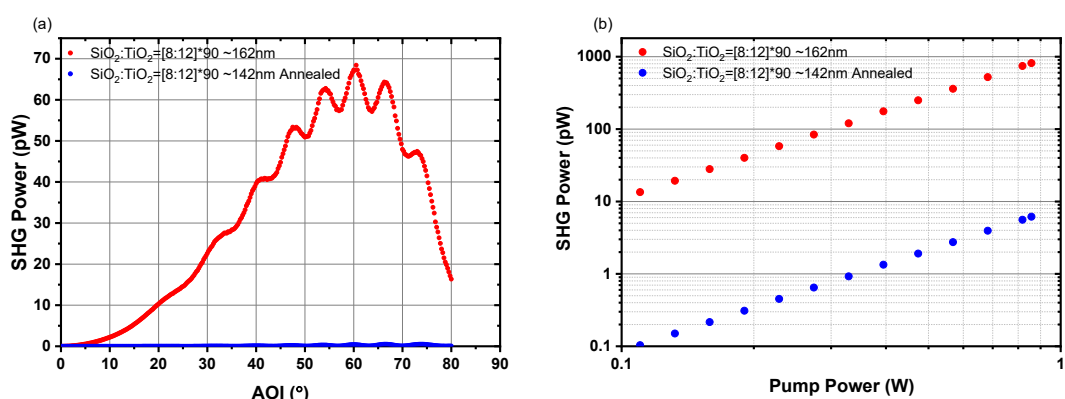

**Figure S6.** Measured SH signals (a) angular and (b) power dependency from $SiO_2:TiO_2$ = [8:12]*90 before and after annealing. The SHG slopes are 1.99(1) and 2.00(1).

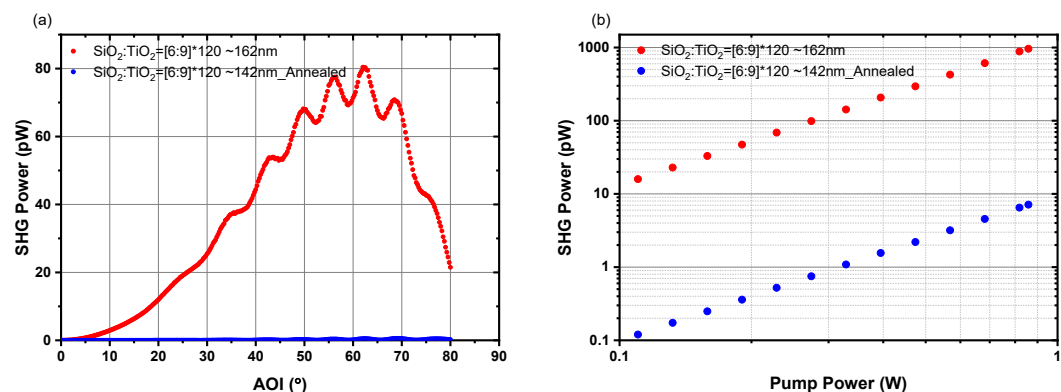

**Figure S7.** Measured SH signals (a) angular and (b) power dependency from $SiO_2:TiO_2$ = [6:9]*120 before and after annealing. The SHG slopes are 2.00(1) and 1.99(1).

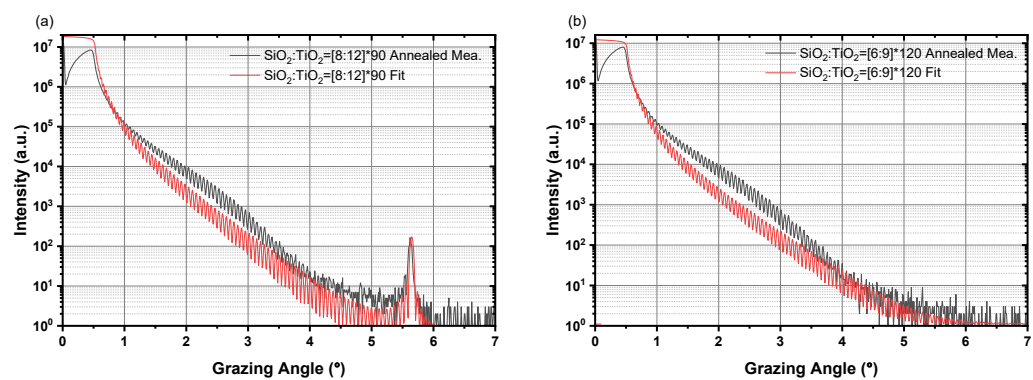

**Figure S8.** XRR measurements and fitting for annealed samples (a) $SiO_2:TiO_2 = [8:12]*90$ and (b) $SiO_2:TiO_2 = [6:9]*120$.

**Table S1.** XRR data of annealed samples.

| Compositions $SiO_2:TiO_2$ | Interface roughness (nm) $SiO_2$ \| $TiO_2$ | Layer thickness (nm) $SiO_2$ \| $TiO_2$ | Target total thickness (nm) | Total thickness (nm) Before annealing | Total thickness (nm) After annealing |
|---|---|---|---|---|---|
| [6:9]*120 | 0.6 \| 0.6 | 0.7 \| 0.5 | 162 | 158.2 | 141.5 |
| [8:12]*90 | 0.7 \| 0.7 | 0.8 \| 0.8 | 162 | 160.5 | 141.3 |